# A New Scoring Method for the Evaluation of Vehicle Road Departure Detection Systems

Dan Shen, Lingxi Li, Stanley Chien, Yaobin Chen,
Transportation Active Safety Institute
Indiana University-Purdue University Indianapolis, USA

Rini Sherony
Collaborative Safety Research Center (CSRC),
Toyota Motor North America, USA

*Abstract*—**Road departure detection systems (RDDSs) for eliminating unintentional road departure collisions have been developed and equipped on some commercial vehicles in recent years. In order to provide a standardized and objective performance evaluation of RDDSs without the affections of systems' complex nature of RDDSs and the design requirements, this paper proposes the development of the scoring method for evaluating vehicle RDDSs. Both flat road edge and vertical road edge are considered in the proposed scoring method, which combines two key variables: 1) the lateral distance of vehicle from road edge when RDW triggers; 2) the lateral distance of vehicle from road edge when RKA triggers. Two main criteria of road departure warning (RDW) and Road Keeping Assistance (RKA) are used to describe the performance of RDDSs.**

## I. Introduction

Vehicle accidents due to road departure is a leading cause of injuries and fatalities on US highways [1]. Approximately 12,000 drivers lose their lives each year due to road departure collisions [1]. Roadside crashes account for about 35 percent of the fatalities on nation's highways [2]. Road departure detection system are becoming increasingly widespread nowadays. Road departure warning (RDW) and road keeping assistance (RKA) systems are two new active safety technologies in RDDSs for dealing with the run-off road issue. Most of the currently developed lane/road departure detection systems are based on the lane marking detections. However, since there are no representative global standards for developing RDDSs, companies equipped their own RDDSs according to different design principles and their complicated nature. Thus, the performance of different RDDSs vary significantly under different environment.

Safety professionals and automobile manufacturers have strived to overcome the driving safety issue and develop a standard vehicle performance evaluation mechanism and strategies for the Advanced Driving Assistance Systems (ADASs). In particular, the relevance of the New Car Assessment Program (NCAP) Lane Departure Warning (LDW) confirmation test to real-world road departure crashes was studied in [3]. The authors in [4] introduced an scoring methodology for the establishment of a standard to rate and compare the performance of Crash Imminent Braking (CIB) systems A set of test scenarios on test track is also developed for the performance evaluation of CIB systems. Reference [5] explored the effectiveness of the scoring approach in [4], which describes a systematic method for CIB systems evaluation with considering both warning and braking features. An approach for situation assessment of emergency braking system for vehicle-to-vehicle scenarios was presented in [6]. To date, there are no representative standards to evaluate and compare the comprehensive performance of different RDDSs except for the assessment of Lane Support System (LSS) developed by Euro-NCAP [7,8]. LSSs can issue waring signal to the driver when driver unintentionally drift out the road lane or when change lane without indication through LDW and assist driver correct the heading of the vehicle if driver fails to take actions via LKA.

This study is to propose a scoring method for the evaluation of RDDSs. Before getting into details, we first need to clarify what is considered as a road departure. Since we assume that there may not be road edge marking on many roads, the road departure is defined here as the vehicle passes the edge between the paved homogenous road surface and unpaved rough roadside [9]. Road departure can be divided into two types, one is the driver intended road departure, and the other is driver unintended road departure. Road Departure Detection Systems are designed mainly for handling driver unintended road departures. Then one of the questions in RDDS is to distinguish if a road departure is intended or unintended by the driver. Since it is hazardous for an RDDS to mitigate a driver intended road departure, RDDS should only take actions when there is a driver unintended road departure. The RDDS may rely on the driver's explicit indication of the intended road departure, such as the use of turn signal and turning the steering wheel. However, it is still difficult to figure out if a road departure is intentional or unintentional on low-speed streets. Therefore, scoring the performance of RDDS in user intended road departures is out of the scope of this study. This paper only gives the scoring method for vehicle testing under the scenarios of driver unintended road departures according to the Euro-NCAP Assessment of Lane Support Systems [7, 8].

The existed scoring approach for evaluating RDDSs defined by Euro-NCAP is only considered the flat road edge such as grass, dirt and gravel [7,8]. The primary goal and main contribution of this work is the development of a comprehensive scoring method for rating and comparing the performance of vehicle road departure detection systems on all types of roads with/without lane markings. This scoring method includes both flat road edge and vertical road edge, and, two key parameters : 1) the lateral distance of test vehicle from road edge when RDW is triggered; 2) the lateral distance of test vehicle from road edge when RKA is triggered, which provide a practical guidance on the development of next generation of intelligent RDDSs.

The remainder of this paper is organized as follows. The criteria and scoring method used by Euro-NCAP for evaluating the lane support system are introduced and discussed in Section II. The proposed scoring method and criteria for evaluating the road departure detection systems are presented in Section III. Section IV describes the score parameters and the aggregation of test scores. Finally, the conclusions and comments on the proposed scoring approach are given in Section V.

## II. Criteria and Scoring in Euro-NCAP

Lane support systems are becoming more and more widespread. Euro-NCAP has acknowledged its safety potential through the Euro NCAP Advanced award process. From 2014, these systems are included in the Safety Assist score. Here we generalize the lane support system in Euro-NCAP report to use the relevant concept in road departure mitigation systems. The successful three-stage working process of current RDDSs on vehicle potential straying road that may cause crashes has been introduced in [9]. At stage 1, the warning signal is triggered when vehicle lateral deviation or potential lane/road departure was detected by vehicle onboard sensors; Then the steering assistance (and possibly braking assistance) input will be applied to avoid lane/road departure if driver fails to take actions at the second stage; Stage 3 can maintain vehicle back onto the center-line of the lane/road using autonomous steering.

Since the Euro-NCAP uses the lane related definitions for scoring the RDDS, the lane related definitions are listed here and will be used for the RDDS scoring method:

1) **Lane Edge** – means the inner side of the lane marking or the road edge.
2) **Emergency Lane Keeping (ELK)** – default ON heading correction that is applied automatically by the vehicle in response to the detection of the vehicle that is about to drift beyond the edge of the road or into oncoming or overtaking traffic in the adjacent lane.
3) **Lane Keeping Assist (LKA)** – heading correction that is applied automatically by the vehicle in response to the detection of the vehicle that is about to drift beyond a delineated edge line of the current travel lane.
4) **Lane Departure Warning (LDW)** – a warning that is provided automatically by the vehicle in response to the vehicle that is about to drift beyond a delineated edge line of the current travel lane.
5) **VUT** – VUT means vehicle under test. It is assumed that the VUT is equipped with a Road Keeping Assist and/or Road Departure Warning system.
6) **Time to Line Crossing (TTLC)** – the remaining time before the VUT crosses the line, if the VUT would continue to travel with the same lateral velocity towards the lane marking.
7) **Distance to Lane Edge (DTLE)** – means the remaining lateral distance (perpendicular to the Lane Edge) between the Lane Edge and the outermost edge of the tire, before the VUT crosses Lane Edge, assuming that the VUT would continue to travel with the same lateral velocity towards it. Shown in Figure 2.1.

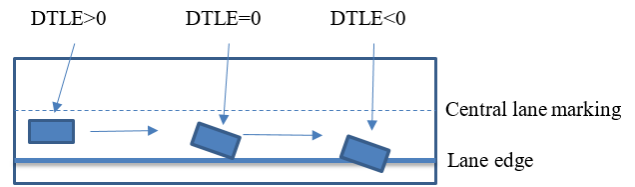

Fig. 2.1. Illustration of distance to lane/road edge (DTLE).

The following part of this section will list the Euro-NCAP rate method for the lane support systems (LSS) which is relevant to this study. There are three scoring criteria applied in the Euro-NCAP report, namely, Human Machine Interface, Lane Keep Assist, and Emergency Lane Keeping [6, 7]. But only two scoring criteria (Human Machine Interface and Lane Keep Assist) are related to our scoring method, and their specific descriptions are listed as below:

1. Human Machine Interface (HMI)

HMI points can be achieved for the following:

- Lane Departure Warning (LDW) (0.25 points)

Any LDW system that issues an audible/ and/or haptic warning before a DTLC of -0.2m is awarded.

2. Lane Keeping Assist (LKA)

LKA points can be achieved for the following:

- For LKA system tests, the assessment criteria used is the Distance to Lane Edge (DTLE).
- The limit value for DTLE for LKA tests is set to -0.3m for testing against lines, meaning that the LKA system must not permit the VUT to cross the inner edge of the lane marking by a distance greater than 0.3m.
- The limit value of DTLE for LKA Road Edge tests is set to -0.1m for testing against the road edge, meaning that

the LKA system only allows the VUT to have a part of the front wheel outside of the road edge.

- The available points per test are awarded based on a pass/fail basis where all tests within the scenario and road marking combination need to be a pass. The points available for the different LKA scenario and road marking combinations are detailed in the table I below:

Table I. The scores for different LKA scenarios

| LKA Scenario | Road Marking | Points |
|---|---|---|
| Road Edge | Road Edge only | 0.25 |
|  | Road Edge with central lane marking | 0.25 |

### III. PROPOSED CRITERIA AND SCORING METHOD

Since the scoring method developed by Euro-NCAP only covers the flat road edge, such as grass, dirt, gravel and so on. The scoring system for the vertical road edges like a concrete divider and the metal guardrail is not considered. Before describing the detailed scoring criteria and the scoring method for both the flat road edge and the vertical road edge. The definition of vehicle dimensional measurements will be discussed. This is because the vehicle dimensional measurements are crucial to the definitions of the outermost edge of the testing vehicle, which will be used in the scoring method.

*Vehicle dimensional measurements:*

Vehicle dimensions shall be represented by a two-dimensional polygon defined by the lateral and longitudinal dimensions relative to the centroid of the vehicle using the standard SAE coordinate system. The corners of the polygon are defined by the lateral and longitudinal locations where the plane of the outside edge of each tire makes contact with the road. This plane is defined by running a perpendicular line from the outer most edge of the tire to the ground at the wheelbase, as illustrated in Figure 3.1. For the flat road edge,

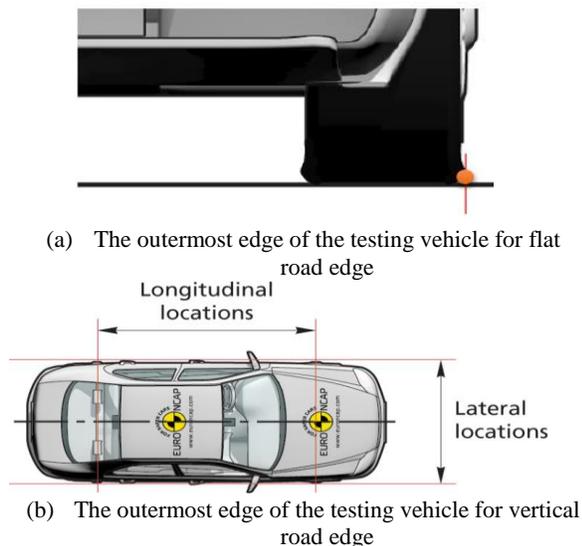

(a) The outermost edge of the testing vehicle for flat road edge

(b) The outermost edge of the testing vehicle for vertical road edge

Fig. 3.1. Vehicle dimensional measurements.

since the testing vehicle still can cross the road edge after road departure. Thus, we define the outermost edge of the testing vehicle as shown in Figure 3.1 (a), which is the outermost edge of the tire. For the vertical road edge like a concrete divider and metal guardrail, since they have heights along the road and the testing vehicle cannot cross them after road departure with a very small angle. Thus, the outermost edge of the testing vehicle should be the outermost edge of the vehicle as Figure 3.1 (b), which is the first collision point.

According to the assessment method from Euro-NCAP, the limitation of our test condition and the definitions of the outermost edge of the testing vehicle, we considered two criteria for scoring the vehicle road departure testing on both the flat road edge and the vertical road edge. Before we go more detailed discussion into the scoring method, the local coordinate setups for scoring vehicle right departure testing and the vehicle left departure testing with both flat road edge and vertical road edge are depicted in Figures 3.2 and 3.3. The convention used for the local coordinate system utilizes the right-handed set with the Z-axis upward, which is the same setup approach introduced in Section 7. Thus, vehicle's distance from the road edge is vehicle's location on the Y-axis.

For the flat road edge, the real grass and surrogate grass will be tested on the test track. Since the testing vehicle can cross the road edge (road departure distance 0m) in the testing. Thus, the vehicle lateral position can be described as follow for further scoring analysis:

- For right road departure, the vehicle lateral position is negative when the vehicle is on the road and positive when the vehicle is off the road edge.
- For left road departure, the vehicle lateral position is positive when the vehicle is on road and negative when the vehicle is off the road edge.

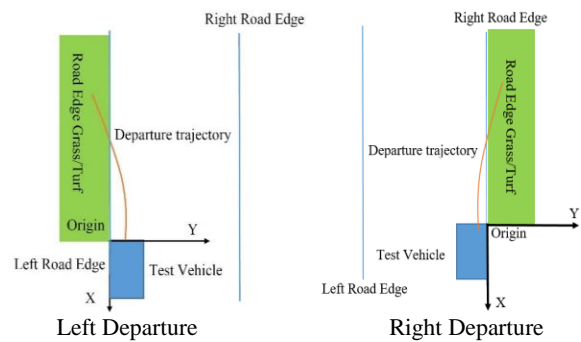

Left Departure      Right Departure

Fig. 3.2. Local coordinate setup for flat road edge.

For the vertical road edge such as metal guardrail and concrete divider, it is very difficult to test them on real test track due to the safety and limitation of the testing facility. Since the testing vehicle cannot cross the road edge (road departure distance 0m) in the testing. Thus, the vehicle lateral position can be described as follow for further scoring analysis:

- For right road departure, vehicle lateral position is always negative when the vehicle is on the road and zero when the vehicle struck the road edge.
- For left road departure, the vehicle lateral position is always positive when the vehicle is on road and zero when the vehicle struck the road edge.

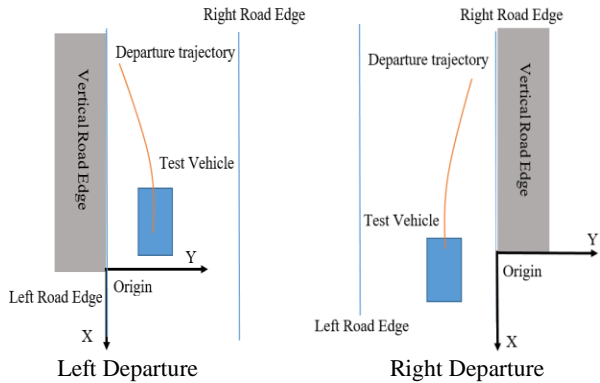

Fig. 3.3. Local coordinate setup for vertical road edge.

Therefore, two scoring criteria are listed below based on the local coordinate system for both the flat road edge and the vertical road edge:

1. Human Machine Interface (RDW of HMI)

- Flat Road Edge

*Right Departure*

HMI points can be accomplished for the Road Departure Warning (RDW). Any RDDS that issues an audible and/or haptic warning before a Distance to Road Edge (DTRE) of 0.2m is awarded. The 0.2m indicates that the position of the outermost edge of the vehicle tire is 0.2m beyond the road edge, meaning that the RDW system must not permit the VUT to cross the road edge line by a distance greater than 0.2m. (0.25 points).

*Left Departure*

HMI points can be accomplished for the Road Departure Warning (RDW). Any RDDS that issues an audible/ and/or haptic warning before a Distance to Road Edge (DTRE) of -0.2m is awarded. The -0.2m indicates that the position of the outermost edge of the vehicle tire is 0.2m beyond the road edge, meaning that the RDW system must not permit the VUT to cross the road edge line by a distance greater than 0.2m. (0.25 points).

- Vertical Road Edge

*Right Departure*

HMI points can be accomplished for the Road Departure Warning (RDW). Any RDDS that issues an audible/ and/or haptic warning before a DTRE of -0.2m is awarded. The -0.2m indicates that the position of the outermost edge of the vehicle is 0.2m before touches the road edge, meaning that the RDW system must not permit the VUT to touch the road edge line within the distance of 0.2m. (0.25 points).

*Left Departure*

HMI/RDW points can be accomplished for the Road Departure Warning (RDW). Any RDDS that issues an audible/ and/or haptic warning before a DTRE of 0.2m is awarded. The 0.2m indicates that the position of the outermost edge of the vehicle is 0.2m before touches the road edge, meaning that the RDW system must not permit the VUT to touch the road edge line within the distance of 0.2m. (0.25 points).

2. Road Keeping Assist (RKA)

- Flat Road Edge

*Right Departure*

For RDDS testes, the assessment criteria used is the Distance to Road Edge (DTRE). The limit value for DTRE for RDDS road edge tests is set to 0.1m for testing against the road edge. The 0.1m indicates that the position of the outermost edge of the vehicle tire is 0.1m beyond the road edge, meaning that the RKA system must not permit the VUT to cross the road edge by a distance greater than 0.1m. (0.25 points)

*Left Departure*

For RDDS testes, the assessment criteria used is the Distance to Road Edge (DTRE). The limit value for DTRE for RDDS road edge tests is set to -0.1m for testing against the road edge. The -0.1m indicates that the position of the outermost edge of the vehicle tire is 0.1m beyond the road edge, meaning that the RKA system must not permit the VUT to cross the road edge by a distance greater than 0.1m. (0.25 points)

- Vertical Road Edge

*Right Departure*

For RDDS testes, the assessment criteria used is the Distance to Road Edge (DTRE). The limit value for DTRE for RDDS road edge tests is set to -0.1m for testing against the road edge. The -0.1m indicates that the position of the outermost edge of the vehicle is 0.1m before touching the road edge, meaning that the RKA system must not permit the VUT to touch the road edge within the distance of 0.1m. (0.25 points).

*Left Departure*

For RDDS testes, the assessment criteria used is the Distance to Road Edge (DTRE). The limit value for DTRE for RDDS road edge tests is set to 0.1m for testing against the road edge. The 0.1m indicates that the position of the outermost edge of the vehicle is 0.1m before touches the road edge, meaning that the RKA system must not permit the VUT to touch the road edge within the distance of 0.1m. (0.25 points).

Therefore, we will develop the scoring method based on the concepts and factors of these two criteria. For better illustration, the demonstration of road environment descriptions for both flat road edge and vertical road edge can be depicted in Figures 3.4 and 3.5, respectively.

- The road edge line is the real road edge and is known as the boundary between the paved road and unpaved roadside.
- The Road Keeping Assist (RKA) fail line is a line indicating that the road departed vehicle must start applying RKA to avoid a crash, which is 0.1m away beyond the flat road edge and 0.1m before striking the vertical road edge.
- The Road Departure Warning (RDW) fail line is another line indicator that the road departed vehicle must start issuing an audible and/or haptic warning to the driver to avoid a crash, which is 0.2m away beyond the flat road edge and 0.2m before striking the vertical road edge.

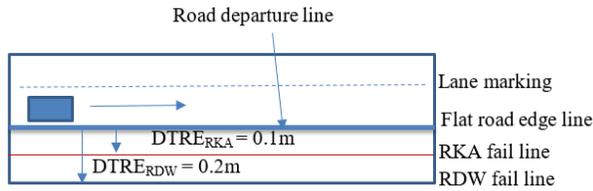

Fig. 3.4. Illustration of various road lines for the flat road edges.

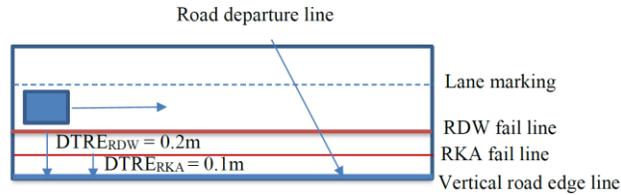

Fig. 3.5. Illustration of various road lines for the vertical road edges.

**Assumptions**
- VUT is on the road at the beginning of the test.
- VUT would continue to travel with the same lateral velocity towards the road edge.

**Nomenclatures**
- $DTRE_{RKA}$ - The distance from the road edge line to the RKA fail line that the VUT is considered to take assistance to avoid road departure.
- $DTRE_{RDW}$ - The distance from the road edge line to the RDW fail line that the VUT is considered to issue an audible and/or haptic warning to the driver to avoid a crash.
- $d_{lateral}$ - The lateral distance of the VUT from the road edge.
- $d_{RKA}$ - The lateral distance of the VUT from the road edge when the VUT starts road keeping assist.
- $d_{RDW}$ - The lateral distance of the VUT from the road edge when the VUT starts road departure warning signal.
- $d_{max-lateral}$ - The maximum lateral distance of the VUT from the road edge during the test

- $DTRE_{RKA}$ and $DTRE_{RDW}$ are defined as 0.1m and 0.2m, respectively, for road edge according to the Euro-NCAP Lane Support System Assessment protocol. Therefore, the scoring method should include and rely on the information of $d_{RKA}$ and $d_{RDW}$.

IV. SCORE PARAMETERS AND SCORE AGGREGATION

In this paper, two parameters can be used to describe the performance of road departure mitigation: warning and steering. Therefore, there are 4 exclusive combinations for these two parameters as shown in Table II.

Table II. Different combinations of warning and steering in RDDS actions.

| Cases | Warning | Steering |
|---|---|---|
| 1 | warning issued | steering issued |
| 2 | warning not issued | steering issued |
| 3 | warning issued | steering not issued |
| 4 | warning not issued | steering not issued |

- In a road departure test, a warning may or may not be issued.
- In a road departure test, a RKA steering may or may not be activated.

The score for each parameter is defined as follows:

$S_w$ = the score for the performance of road departure warning

$S_s$ = the score for the performance of road departure steering

$S_w$ and $S_s$ should have a positive value if the respective warning and steering help road safety. There could be a big discussion of what help road safety means. In this scoring method, we only consider clear-cut cases. $S_w$ and $S_s$ should be 0 if the respective warning and steering are not activated. Let us set the maximum score as 0.25 each to $S_w$ and $S_s$ for convenience.

The descriptions for the determination of $S_w$ and $S_s$ values are in later sections. We can give a combined score, $S$, for the RDMS performance of all warning and steering as follows.

$$S = w_w \times S_w + w_s \times S_s \qquad (1)$$

Where $w_w$ and $w_s$ are the weighting factors of the score for warning $S_w$ and the score for steering $S_s$, respectively. Then the next question is how to assign values to each of these two weighting factors. Here we ignore this factor and assign both $w_w$ and $w_s$ equal to 1.

Then a natural question will be raised that hot to assign the score for road departure detection systems. Based on the assessment protocol from Euro-NCAP, the following two constraints in each test scenario determine the performance of RDDS:

1. $d_{RKA}$, the lateral distance of the VUT from the road edge when the VUT starts road keeping assist.

2. $d_{RDW}$, the lateral distance of the VUT from the road edge when the VUT starts road departure warning.

As depicted in Figures 4.1 and 4.2, we introduce the scoring method for vehicle right departure testing with both flat road edge and vertical road edge as follows. The scoring method for left departure can be easily determined by reversing the sign of right departure cases according to local coordinate setup methods in Section III.

### 1. Flat Road Edge

*Right departure cases*

For RDW score in vehicle right departure, if RDMS of a VUT can trigger warning signal before VUT touching the RDW fail line ($d_{RDW} < 0.2$m), the RDW score will be awarded as 0.25, and if not function at all ($d_{RDW} > 0.2$m), the RDW score is 0.

For RKA score in right departure, if RDMS of a VUT can apply auto-steering before VUT touching the RKA fail line ($d_{RKA} < 0.1$m), the RKA score will be awarded as 0.25, and if not function at all ($d_{RKA} > 0.1$m), the RKA score is 0.

*Left departure cases*

For RDW score in a vehicle left departure, if RDMS of a VUT can trigger warning signal before VUT touching the RDW fail line ($d_{RDW} > -0.2$m), the RDW score will be awarded as 0.25, and if not function at all ($d_{RDW} < -0.2$m), the RDW score is 0.

For RKA score in left departure, if RDMS of a VUT can apply auto-steering before VUT touching the RKA fail line ($d_{RKA} > -0.1$m), the RKA score will be awarded as 0.25, and if not function at all ($d_{RKA} < -0.1$m), the RKA score is 0.

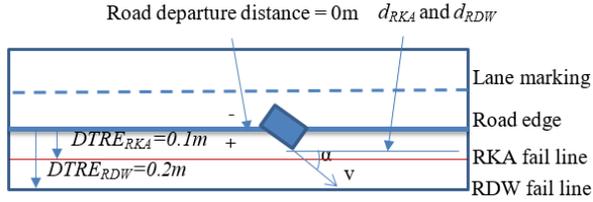

Fig. 4.1. VUT departs from the flat road edge with velocity v and departure angle α.

### 2. Vertical Road Edge

*Right departure cases*

For RDW score in vehicle right departure side, if RDMS of a VUT can trigger warning signal before VUT touching the RDW fail line ($d_{RDW} < -0.2$m), the RDW score will be awarded as 0.25, and if not function at all ($d_{RDW} > -0.2$m), the RDW score is 0.

For RKA score in vehicle right departure side, if RDMS of a VUT can apply auto-steering before VUT touching the RKA fail line ($d_{RKA} < -0.1$m), the RKA score will be awarded as 0.25, and if not function at all ($d_{RKA} > -0.1$m), the RKA score is 0.

*Left departure cases*

For RDW score in a vehicle left departure side, if RDMS of a VUT can trigger warning signal before VUT touching the RDW fail line ($d_{RDW} > 0.2$m), the RDW score will be awarded as 0.25, and if not function at all ($d_{RDW} < 0.2$m), the RDW score is 0.

For RKA score in a vehicle left departure side, if RDMS of a VUT can apply auto-steering before VUT touching the RKA fail line ($d_{RKA} > 0.1$m), the RKA score will be awarded as 0.25, and if not function at all ($d_{RKA} < 0.1$m), the RKA score is 0.

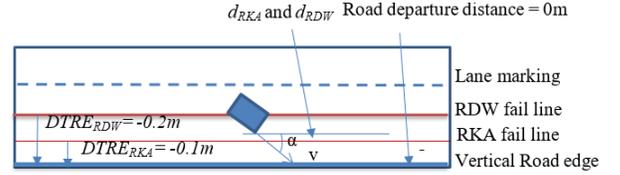

Fig. 4.2. VUT departs from the vertical road edge with velocity v and departure angle α.

The total score in points for each test run is the sum of the RDW score and RKA score for the following:

| LSS function | Points |
|---|---|
| RDW | 0.25 |
| RKA | 0.25 |
| Total | 0.5 |

Therefore, the total performance score of one test run of a specific road departure scenario can be

$$S^i = S^i_w + S^i_s \qquad (2)$$

Where $S^i_w$ is the score for the RDW and $S^i_s$ is the score for the LKA in test scenario i. The full score of $S^i_w$ and $S^i_s$ are both 0.25. Thus, the full score of the RDDS is 0.5.

If there are multiple test runs (m) of a specific road departure scenario i, their performance score $S^i$ is the average of the scores of those tests.

$$S^i = \frac{1}{m} \times \sum_1^m S^i_m \qquad (3)$$

Where m is the number of all test runs in one test scenario.

If we get the final performance score, $S_{comprehensive}$, for all *n* road departure test scenarios, they can be aggregated based on the

$$S_{comprehensive} = \sum_{i=1}^n w^i S^i, \text{ and } \sum_{i=1}^n w^i = 1 \qquad (4)$$

Where *n* is the number of all test scenarios and $w^i$ can be assigned equal weights for each road departure scenario (1/n).

## V. CONCLUSIONS

This work focused on the scoring method for evaluating the performance of vehicle road-departure detection systems, which is mainly based on road departure warning and road keeping assistance two criteria. The overall mechanism of proposed scoring method was introduced. Two key variables for defining and describing the performance of RDDSs were determined according to the scoring method of lane support system defined by Euro-NCAP. Two key parameters are the lateral distance of test vehicle from road edge when road departure warning is triggered, the lateral distance of test vehicle from road edge when road keeping assistance is triggered. In the meantime, both the flat road edge and vertical road edge are considered in the proposed scoring approach. In the future, some other research directions may be explored further such as assigning the scenario weighting factors based on the occurrence frequency and how to aggregate the comprehensive test score with different design principle.